\documentclass{amsart}
\usepackage{amssymb}
\usepackage{latexsym}
\usepackage{amsmath}
\usepackage{euscript}
\usepackage{graphics}
\usepackage[all]{xy}

\newcommand{\be}{\begin{equation}}
\newcommand{\ee}{\end{equation}}
\newcommand{\ba}{\begin{eqnarray}}
\newcommand{\ea}{\end{eqnarray}}
\newcommand{\baa}{\begin{eqnarray*}}
\newcommand{\eaa}{\end{eqnarray*}}
\newcommand{\bb}{\begin{thebibliography}}
\newcommand{\eb}{\end{thebibliography}}

\newcommand{\bi}[1]{\bibitem{#1}}
\newcommand{\lab}[1]{\label{#1}}
\newcommand{\re}[1]{(\ref{#1})}

\newcounter{my}
\newcommand{\he}%
   {\stepcounter{equation}\setcounter{my}%
   {\value{equation}}\setcounter{equation}0%
   }%
\newcommand{\she}%
   {\setcounter{equation}{\value{my}}%
    }%

\renewcommand\t{\tilde}

\newcommand\ve{\varepsilon}

\newtheorem{pr}{Proposition}

\theoremstyle{definition}

\numberwithin{equation}{section}

\begin{document}


\title{Tridiagonalization and the Heun equation}

\author{F.Alberto Gr\"unbaum}
\author{Luc Vinet}
\author{Alexei Zhedanov}

\address {Department of Mathematics, 
University of California, 
Berkeley CA 94720}

\address{Centre de recherches math\'ematiques
Universit\'e de Montr\'eal, P.O. Box 6128, Centre-ville Station,
Montr\'eal (Qu\'ebec), H3C 3J7}

\address{Institute for Physics and Technology\\
R.Luxemburg str. 72 \\
83114 Donetsk, Ukraine \\}

\begin{abstract}
It is shown that the tridiagonalization of the hypergeometric operator $L$ yields the generic Heun operator $M$. The algebra generated by the operators $L,M$ and $Z=[L,M]$ is quadratic and a one-parameter generalization of the Racah algebra. The new Racah-Heun orthogonal polynomials are introduced as overlap coefficients between the  eigenfunctions of the operators $L$ and $M$. An interpretation in terms of the Racah problem for $su(1,1)$ algebras and separation of variables in a superintegrable system are discussed.
\end{abstract}


\maketitle

\section{Introduction}
\setcounter{equation}{0} 
The purpose of this paper is to show how the Heun differential operator can be obtained from the hypergeometric operator by a procedure called tridiagonalization. This will offer a cogent algebraic framework for the Heun equation and its degenerate cases.The analysis presented here is in part motivated by important "classical" results in data analysis. Typically, signals can only be observed within certain frequency windows and during restricted time intervals. Such limiting is described by a time-frequency integral localization operator that act on the signal function. Strikingly, there is a differential operator that commutes with that integral operator. This result can can be traced back to certain references in Ince \cite{In}. In a remarkable series of papers, \cite{SlPo}, \cite{LaPo}, \cite{LaPo2}, (for reviews see \cite{Sl2}, \cite{Sl3}), Landau, Pollak and Slepian have independently rediscovered and made used of that fact in the study and the applications of the integral operator eigenfunctions. This has also appeared in Mehta's work on random matrices \cite{Me}; further analogous operators commuting with different integral operators have played a critical role in the study of level-spacing distributions of Hermitian matrices of large order at the edge of the spectrum by Tracy and Widom\cite{TrW1}, \cite{TrW2}. 
\\
Interestingly the commuting operator is of the Heun type. The questions as to why there exists such commuting operators and why are they of the Heun type have been intriguing. One of us has been exploring this matter for quite sometime and broadly documented the apparent relation between the existence of the commuting operator, bispectrality and KdV and Virasoro potentials (see in particular \cite{DG}, \cite{G}). In a few recent studies \cite{IK1}, \cite{IK2}, \cite{GIVZ}, it has been shown how more involved operators can be obtained from simpler hypergeometric operators by combining those with the operator corresponding to multiplication by the variable. This method has been referred to as tridiagonalization and the operators thus obtained bear a structural resemblance with the commuting operators we have been discussing. This brought up the question if the Heun operator could be obtained by tridiagonalization. This will be answered in the affirmative. While the results do not bring the last word on the intriguing questions mentioned above, they put them in an interesting light.
\\
The outline of the paper is the following: in Section 2 the most general operator M that can be obtained by tridiagonalization from a hypergeometric operator L will be presented. It will be recalled that in a special case L and M realize the Racah algebra. It will be seen in Section 3 that the most general tridiagonalization leads to M being the Heun operator and that it can thus be algebraically characterized as being 3-diagonal in the basis of Jacobi polynomials. The Racah-Heun algebra encoding this will be introduced.  In this context, it will prove possible to bring about the new Racah-Heun orthogonal polynomials as the expansion coefficients of the polynomial solutions of the Heun equation in the basis of the Jacobi polynomials. In Section 4, it will be shown how the Racah-Heun algebra is realized in the framework of the Racah problem for the $su(1,1)$ algebra. This will allow to make the connection with superintegrable models and to clarify the relation between the Racah-Heun algebra and separation of variables in elliptic coordinates. The finite-dimensional representations of the Racah-Heun algebra are discussed in Section 5. The algebraic treatment of the confluent and doubly confluent Heun equation is the object of Section 6 and obtained by specialization of the Laguerre and Hermite operators. In the former case, one identifies the associated Hahn-Heun algebra. The paper closes with a short conclusion.
\section{Tridiagonalization}

Tridiagonalization is a procedure which allows to construct new tri-diagonal operators out of two operators $X$ and $L$ where X is multiplication by the variable $x$:
\be X f(x) = x f(x) \lab{X_def} \ee
and $L$ is a second-order differential or difference operator. It is assumed that there exist orthogonal polynomials $P_n(x) = x^n + O(x^{n-1})$ which are eigensolutions of the eigenvalue problem
\be
L P_n(x) = \lambda_n \: P_n(x) \lab{L_P}. \ee 
Because $P_n(x)$ are monic orthogonal polynomials, they satisfy the 3-term recurrence relation
\be
P_{n+1}(x) + b_n P_n(x) + u_n P_{n-1}(x) = x P_n(x) \lab{3_term_P}. \ee
This means that the polynomials $P_n(x)$ are bispectral: they satisfy simultaneously two eigenvalue problems: \re{L_P} and \re{3_term_P}.

Note that the 3-term recurrence relation \re{3_term_P} can be presented in the form
\be
J \pi(x) = x \pi(x), \lab{J_pi_pi} \ee
where $J$ is the Jacobi (i.e. tridiagonal) matrix
\[
J =
 \begin{pmatrix}
  b_{0} & 1 & 0 &    \\
  u_{1} & b_{1} & 1 & 0  \\
   0  &  u_2 & b_2 & 1    \\
   &   &  \ddots &    \ddots  \\
      \end{pmatrix},
\] 
and $\pi(x)$ is a vector with components $\{P_0(x), P_1(x), P_2(x),\dots\}$.

As is well known, if $u_n>0$ and $b_n$ are real then the polynomials $P_n(x)$ are orthogonal with respect to some positive measure $d \sigma(x)$ on the real axis, 
\be
\int_{a}^b P_n(x) P_m(x) d \sigma(x) = h_n \: \delta_{nm}, \lab{ort_P}, \ee 
 where the normalization constant is
\be
h_n = u_1 u_2 \dots u_n \lab{h_def}. \ee
Let us consider the operator
\be
M= \tau_1 X L + \tau_2 L X + \tau_3 X + \tau_4 L \lab{M_def}. \ee
It is clear that due to relations \re{L_P} and \re{3_term_P}, the operator $M$ will be tridiagonal on the polynomial basis $P_n(x)$:
\be
M P_n(x) = \xi_{n+1} P_{n+1}(x) + \eta_n P_n(x) + \zeta_{n} u_n P_{n-1}(x), \lab{M_3-diag} \ee
where 
\ba
&&\xi_{n} = \tau_1 \lambda_{n-1} + \tau_2 \lambda_n + \tau_3, \quad \zeta_{n} = \tau_1 \lambda_{n} + \tau_2 \lambda_{n-1} + \tau_3, \nonumber \\&& \eta_n = (\tau_1 + \tau_2)b_n + \tau_3 b_n + \tau_4 \lambda_n  \lab{xez}. \ea
It is sometimes more convenient to use the orthonormal polynomials $\chi_n(x)$. They are defined as
\be 
\chi_n(x) =\frac{P_n(x)}{\sqrt{h_n}} \lab{chi_P} \ee 
and satisfy the 3-term recurrence relation 
\be
a_{n+1} \chi_{n+1}(x) + b_n \chi_n(x) + a_n \chi_{n-1}(x) = x \chi_n(x) \lab{rec_chi} \ee
with the Hermitian Jacobi matrix $\tilde J$ now reading
\[
\tilde J =
 \begin{pmatrix}
  b_{0} & a_1 & 0 &    \\
  a_{1} & b_{1} & a_2 & 0  \\
   0  &  a_2 & b_2 & a_3    \\
   &   &  \ddots &    \ddots  \\
      \end{pmatrix}.
\] 
Let us focus now on the special case where $\tau_4 = 0$. It was shown in \cite{GIVZ}  that if $L$ is the hypergeometric operator corresponding to the Jacobi polynomials, then (under a minor restriction upon coefficients $\tau_i$ in \re{M_def}) the operator $M$ generated by tridiagonalization as in  \re{M_3-diag} with $\tau_4 = 0$, is again hypergeometric up to a change of the independent variable. Together, the operators $L$ and $M$ form in this case the quadratic Racah algebra. The latter is generated by 2 operators $A_1 = L, A_2 = M$ and their commutator $A_3$. These operators satisfy the commutation relations \cite{GIVZ}  
\begin{align}
\label{W_def}
 \begin{aligned}
 {}
 [A_1,A_2]&=A_3,
 \\
[A_2,A_3]&=\alpha_1\{A_1,A_2\} + \alpha_2 A_2^2 + \gamma_1 A_1 + \delta A_2 + \epsilon_1,
\\
[A_3,A_1]&=\alpha_2\{A_1,A_2\} + \alpha_1 A_1^2 + \gamma_2 A_2 + \delta A_1+ \epsilon_2,
 \end{aligned}
 \end{align}
where $\alpha_1, \alpha_2, \gamma_1, \gamma_2, \delta,\epsilon_1,\epsilon_2$ are real structure constants.

\section{Racah-Heun algebra}
\setcounter{equation}{0}
Consider again the case of the hypergeometric operator defined by
\be
L = x(1-x) \partial_x^2 + (\nu_1 x + \nu_2) \partial_x \lab{L_hyp} \ee
with arbitrary constants $\nu_1, \nu_2$. 

This operator has the Jacobi polynomials
\be
P_n^{(\omega_1, \omega_2)}(x) = \frac{(-1)^n (\omega_1+1)_n}{(\omega_1 +\omega_2 +n+1)_n} \: {_2}F_1 \left(   {-n, n +\omega_1 +\omega_2+1 \atop \omega_1+1} ; x \right) \lab{Jac_P} \ee
as eigensolutions
\be
L P_n^{(\omega_1, \omega_2)}(x) =\lambda_n P_n^{(\omega_1, \omega_2)}(x) \lab{L_Jac}.\ee
Here 
\be
\omega_1 = \nu_2 -1, \; \omega_2 = -1-\nu_1-\nu_2, \; \lambda_n = -n(n-\nu_1-1) \lab{par_Jac}, \ee
\be{_2}F_1 \left(   {a, b\atop c} ; x \right)\ee 
is the hypergeometric series and the standard notation 
\be
(a)_n = a(a+1) \dots (a+n-1) \lab{Pochh} \ee
for the Pochhammer symbol is adopted.

The Jacobi polynomials obey the 3-term recurrence relation \re{3_term_P} 
with coefficients  
\begin{equation}
\label{rec_u_Jac}
u_n =\frac{n(n+\omega_1)(n+\omega_2)(n+\omega_1+\omega_2)}{(2n+\omega_1+\omega_2-1)(2n+\omega_1+\omega_2)^2(2n+\omega_1+\omega_2+1)}, 
\end{equation}
and
\begin{equation}
\label{rec_b_Jac}
b_n = \frac{1}{2} + \frac{\omega_1^2-\omega_2^2}{4} \left(\frac{1}{\omega_1+\omega_2+2n} -  \frac{1}{\omega_1+\omega_2+2n+2} \right). 
\end{equation}
Moreover, these polynomials satisfy the structure relation   \cite{KLS} (we omit the dependence on the parameters $\omega_1, \omega_2$ for brevity)
\be
x(x-1) P_n'(x) = n P_{n+1}(x) + G_n P_n(x) + E_n P_{n-1}(x) \lab{str_Jac} \ee
with
\be
G_n = -\frac{n(\omega_1-\omega_2)(\omega_1+\omega_2+n+1)}{(\omega_1+\omega_2+2n)(\omega_1+\omega_2+2n+2)} \lab{str_G} \ee
and
\be
E_n = - \frac{n(\omega_1+n)(\omega_2+n)(\omega_1+\omega_2+n)(\omega_1+\omega_2+n+1)}{(\omega_1 + \omega_2+2n)^2(\omega_1 + \omega_2+2n+1)(\omega_1 + \omega_2+2n-1)} \lab{E_str}. \ee

Introduce the operator $M$ as defined by formula \re{M_def}. In what follows we shall call $M$ the companion operator of $L$

We already mentioned  that when $\tau_4=0$ the operators $A_1=L$ and $A_2=M$ form the Racah algebra \re{W_def}.

Consider now the more general case with $\tau_4 \ne 0$. In what follows we will assume that
\be
\tau_1 + \tau_2 =1 \lab{tt_norm}. \ee
Indeed, when $\tau_2 \ne -\tau_1$, condition \re{tt_norm} corresponds to a normalization of the coefficients $\nu_i, i=1,2,3,4$. The exceptional case $\tau_2 = -\tau_1$ leads to an operator $M$ of the first order with respect to the derivative $\partial_x$. We will not consider this degenerate case here. 

Assuming condition \re{tt_norm}, the companion operator $M$ can be presented as follows:
\be
M= x(1-x)(x+\tau_4) \partial_x^2 + (\rho_2 x^2 + \rho_1 x + \rho_0) \partial_x + r_1 x + r_0 \lab{M_Heun} \ee
where
\be
\rho_2 = \nu_1 + 2 \tau_1-2, \; \rho_1 = 2 - 2\tau_1 + \nu_2 + \tau_4 \nu_1, \; \rho_0=\tau_4 \nu_2 \lab{rhos} \ee
and
\be
r_1=\tau_3 -\tau_1 \nu_1 + \nu_1, \quad r_0 = \nu_2(1-\tau_1). \lab{rs} \ee
Consider the eigenvalue equation
\be
M \psi(x) = \lambda \psi(x). \lab{M_psi} \ee
This equation can be written in the following form
\be
\psi''(x) + \left(\frac{\gamma}{x} + \frac{\delta}{x-1} + \frac{\ve}{x-d} \right) \psi'(x) +\frac{\alpha \beta x -q}{x(x-1)(x-d)} \psi(x) =0 \lab{red_Heun} \ee
where
\ba
&&\gamma = \nu_2, \; \delta = -\nu_1-\nu_2, \; \ve = 2\tau_2, \; d = -\tau_4 \nonumber \\ 
&&  \alpha \beta = -\tau_3 - \nu_1 \tau_2, \; q = \lambda - \tau_2 \nu_2 .   \lab{gde} \ea
Equation \re{red_Heun} is the generic Heun equation, i.e. the linear second-order differential equation with 4 regular singular points at $0,1,d,\infty$ (see, e.g. \cite{Ronveaux} for details).

There is an additional condition on the parameters of the Heun equation \cite{Ronveaux}
\be
\alpha+\beta-\gamma -\delta-\ve +1 =0 \lab{reg_cond} \ee  
which is needed in order to provide regularity of the singular point $x=\infty$.

We thus see that the most general tri-diagonalzition  \re{M_def} of the hypergeometric operator leads to the generic Heun equation for the eigenvalue problem of the companion operator $M$. Note that condition \re{reg_cond} does not impose restrictions on the parameters $\tau_i$ because the parameters $\alpha,\beta$  only intervene in \re{gde} via the the product $\alpha \beta$ in \re{gde} and can hence always be chosen so that \re{reg_cond} is satisfied.

It follows immediately from \re{M_def} that the Heun operator $M$ is tridiagonal with respect to the basis of the Jacobi polynomials $P_n^{(\omega_1, \omega_2)}(x)$:
\be
M P_n^{(\omega_1, \omega_2)}(x) =  \xi_{n+1} P_{n+1}^{(\omega_1, \omega_2)}(x) + \eta_n P_n^{(\omega_1, \omega_2)}(x) + \zeta_n P_{n-1}^{(\omega_1, \omega_2)}(x) \lab{M_tri_P} \ee
The inverse statement is also true and can be considered as a useful characterization of the Heun operator $M$.

\begin{pr}
Assume that $M$ is a linear differential operator of second order which preserves the space of polynomials and which is 3-diagonal on the Jacobi polynomials $P_n^{(\omega_1, \omega_2)}(x)$, i.e. it satisfies property \re{M_tri_P}. Assume moreover that for any choice of the parameters of the operator $M$, there exist corresponding  parameters $\omega_1, \omega_2$ of the Jacobi polynomials. Then the operator $M$ is the generic Heun operator and can be presented in the form \re{M_def}.
\end{pr}
{\it Proof}. We use ideas developed in \cite{IK1}, \cite{IK2}. From \re{M_tri_P} and from the property that $M$ is a second-order operator that maps polynomials into polynomials, it follows that $M$ should have the expression
\be
M = \pi_3(x) \partial_x^2 + \pi_2(x) \partial_x^2 + \pi_1(x) \lab{M_gen} \ee
where $\pi_i(x)$ is a polynomial of degree $\le i$. 

Relation \re{M_tri_P} now reads
 \be
\pi_3(x) P_n''(x) + \pi_2(x) P_n'(x) + \pi_1(x) P_n(x) = \xi_{n+1} P_{n+1}(x) + \eta_n P_n(x) + 
\zeta_n P_{n-1}(x) \lab{M_tri_P2}. \ee
Using the differential equation \re{L_Jac} we can eliminate the second derivative term in \re{M_tri_P2}. Using then the structure relation, we can get rid of the first derivative term in the remaining expression. Finally we obtain
\ba
&&R(x) \left(n P_{n+1} + G_n P_n(x) + E_n P_{n-1}(x)\right) + \pi_1(x) P_n(x)= \nonumber \\
&&\xi_{n+1} P_{n+1}(x) + \eta_n P_n(x) + \zeta_n P_{n-1}(x),\lab{M_tri_P3} \ea
where
\be
R(x) = \frac{(\nu_1 x + \nu_2) \pi_3(x) +x(x-1) \pi_2(x)}{x^2(x-1)^2} \lab{R_def}. \ee
The rhs of \re{M_tri_P3} is a polynomial. Hence the rational function $R(x)$ 
should be a polynomial too. Comparing the degrees of the numerator and the denominator of $R(x)$ we see that in fact $R(x) = const$. This means that necessarily
\be
\pi_3(x) = \rho_1 x(x-1)(x-a) \lab{pi_3_nec} \ee
with some constants $\rho_1, a$, and moreover that
\be
\pi_2(x) + \rho_1 (\nu_1 x + \nu_2) = \rho_2 x(x-1) \lab{pi_2_nec} \ee
with another constant $\rho_2$. Relation \re{pi_2_nec} determines the second degree polynomial $\pi_2(x)$. The remaining polynomial $\pi_1(x)$ can be an arbitrary linear function. Comparing with \re{M_Heun} we see that the operator $M$ is indeed the generic Heun operator.

Moreover, assume that $\alpha, \beta, \gamma, \delta, \ve, q$ are parameters of the Heun equation \re{red_Heun}. Then from \re{gde} we recover the parameters $\omega_1, \omega_2$ of the corresponding Jacobi polynomials:
\be
\omega_1 = \gamma-1, \quad \omega_2 = \delta-1 \lab{Jac_H_par} \ee
Thus the parameters  of the Jacobi polynomials depend only on 2 parameters $\gamma$ and $\delta$ of the Heun equation.

We have thus arrived at a purely algebraic definition of the Heun operator as an operator which is 3-diagonal with respect to Jacobi polynomials. Equivalently, this 3-diagonal property means that the Heun operator can always be presented in the form  \re{M_def} with appropriate constants $\tau_1, \tau_2, \tau_3, \tau_4$.

Consider now the algebra behind the operators $L$ and $M$. Introducing the commutator $Z=[L,M]$ we arrive at the following quadratic algebra with relations
\begin{align}
\label{Heun_alg}
 \begin{aligned}
 {}
 [L,M]&=Z,
 \\
[M,Z]&=\alpha_1\{L,M\} + \alpha_2 M^2 + \gamma_1 L + \delta M + \kappa L^2 + \epsilon_1,
\\
[Z,L]&=\alpha_2\{L,M\} + \alpha_1 L^2 + \gamma_2 M + \delta L+ \epsilon_2,
 \end{aligned}
 \end{align}
where
\be
\kappa = 6 \tau_4 (\tau_4 +1) \lab{kappa} \ee
and where the the remaining structure constants are given by
\ba
&&\alpha_1 = -4 \tau_4 -2, \; \alpha_2= 2,\nonumber \\
&& \gamma_1 = 4 (\tau_1\tau_2 - \tau_3 \tau_4) +\tau_4 (\nu_1+2)(2 \nu_2 - \tau_4 \nu_1)  + \nu_2 (2-\nu_2), \nonumber \\
&&\gamma_2 = -2 \nu_1 - \nu_1^2, \; \delta = -2\tau_3 +(\nu_1+2)(\tau_4 \nu_1 - \nu_2), \nonumber \\ 
&&\ve_1 =2 \tau_1 \tau_2 \nu_2 (\nu_1+\nu_2) + \tau_3 \nu_2 \left(\tau_4 (\nu_1+2) +2-\nu_2 \right), \; \ve_2 = -\tau_3 \nu_2 (\nu_1+2)
\lab{str_RH}. \ea

This algebra differs from the Racah algebra only by the term $\kappa L^2$. Despite this "small" difference this algebra lies beyond the class of algebras describing tridiagonal pairs \cite{Ter}. The Racah algebra appears only for two special values of the parameter $\tau_4$: either $\tau_4=0$ or $\tau_4=-1$. In both cases the companion operator $M$ becomes the ordinary hypergeometric operator under a simple change of the independent variable (see \cite{GIVZ} for details). For all other choices of the parameter $\tau_4$ we obtain an operator $M$ that leads to the generic Heun equation \re{red_Heun}. It is hence natural to refer to the algebra \re{Heun_alg} as the Racah-Heun algebra.

The Casimir operator $Q$ of the Racah-Heun algebra which commutes with the operators $L$ and $M$ can be presented in the form
\be
Q= Q_0 + Q_1 , \lab{Cas_RH} \ee
where $Q_0$ is the Casimir operator of the Racah algebra \cite{GZ_preprint}, \cite{GZ_6j}, \cite{GLZ_Annals} with $\kappa=0$:
\ba
&&Q_0 = Z^2 + \alpha_1 \{ L^2, M \} + \alpha_2 \{ M^2, L \} + (\alpha_1^2 +\gamma_1) L^2 + (\alpha_2^2 +\gamma_2) M^2 + \\ \nonumber
&& (\delta + \alpha_1 \alpha_2)\{ L,M \} + (\alpha_1 \delta + 2 \epsilon_1)L + (\alpha_2 \delta + 2 \epsilon_2) M \lab{Q0} \ea
and where $Q_1$ is an additional part proportional to the "perturbation" parameter $\kappa$:
\be
Q_1 = \frac{\kappa}{3} \left( 2 L^3 - \alpha_2 L^2 - \gamma_2 L \right) \lab{Q1}. \ee
For the given realization of the Racah-Heun algebra in terms of the differential operators $L$ and $M$ the Casimir operator reduces to the constant
\be
Q =q = -2 \nu_2 \tau_1 \tau_2 (2+\nu_1 )(\nu_1+\nu_2) - \nu_2 \tau_3 \left((2-\nu_2)\tau_3 +2(2+\nu_1)(\tau_4+1) \right) \lab{Q=q}. \ee

This algebra has the following features:

(i) there exists a basis $p_n$ such that the operator $L$ is diagonal while the operator $M$ is tri-diagonal:
\be
L p_n = \lambda_n p_n , \quad M p_n = \xi_{n+1} p_{n+1} + \eta_n p_n + \zeta_{n} u_n p_{n-1} \lab{3-diag_H_alg} \ee
with $\lambda_n$ a quadratic function in $n$. The matrix coefficients $\xi_n, \eta_n, \zeta_n$ can be obtained directly from the representations of the algebra \re{Heun_alg}. 

\vspace{5mm}

(ii) there exists a realization of the Racah-Heun algebra in terms of second-order differential operators $L$ and $M$. In this case the operator $L$ is the standard hypergeometric operator while $M$  is the generic Heun operator.

\vspace{5mm}

For our concrete realization of the operators $L$ and $M$ we have
\ba
\lambda_n = n(\nu_1 +1 -n) \lab{lambda_Jac} \ea
and the coefficients $\xi_n, \eta_n, \zeta_n$ are given by \re{xez} where $b_n,u_n$ are the recurrence coefficients of the monic Jacobi polynomials \cite{GIVZ}.

The basis $p_n$ consists of hypergeometric functions (Jacobi polynomials in the finite-dimensional case). As we showed, this property means that the Heun operator $M$ is 3-diagonal in this basis. This leads to the property that the expansion coefficients of the Heun functions with respect to the hypergeometric functions satisfy a 3-term recurrence relation. This property is well known (see, e.g. \cite{Erd}, \cite{KM}). We have here a simple algebraic explanation of this phenomenon.

We shall consider now a special finite-dimensional representation of the Racah-Heun algebra. We start with the concrete realization of the operators $L$ and $M$ by the differential operators \re{L_hyp}, \re{M_Heun}. Both operators act on the space of polynomials. The operator $L$ preserves the degree of any polynomial while the operator $M$ increases this degree by 1. Hence in order to obtain a finite-dimensional representation of the Racah-Heun algebra, we need a truncation condition so that the action of the operator $M$ on the monomial $x^N$ gives only a polynomial of degree $N$. We shall then have a $N+1$-dimensional space of polynomials of degrees $0,1,\dots, N$ which is invariant under the action of the operators $L$ and $M$ and hence the Racah-Heun algebra will have finite-dimensional representation on that space.

In order to find this condition let us consider the action of the operator $M$ on the monomials $x^n,\; n=0,1,2,\dots$. We have from \re{M_Heun}
\be
M x^n = (-n(n-1) + \rho_2 n + r_1) x^{n+1} + O(x^n). \lab{M_x_n} \ee
Hence the truncation condition is
\be
-N(N-1) + \rho_2 N + r_1=0. \lab{tr_M} \ee 
Under condition \re{tr_M}, the operators $L$ and $M$ both act on the space of polynomials of degree $\le N$.

We can therefore find polynomial eigensolutions of the operator $M$ on this space
\be M Q_n(x) = \tilde \lambda_n \: Q_n(x), \quad n=0,1,2,\dots,N \lab{M_Q_lam} \ee
where $Q_n(x)$ are some polynomials of degree $\le N$ and where $\t \lambda_n$ are the corresponding eigenvalues. Note that the degree of the polynomial $Q_n(x)$ is NOT equal to $n$. Instead, all polynomials $Q_n(x)$ have, in general, the maximal degree $N$. 

The polynomials $Q_n(x)$ are not orthogonal polynomials. They are called the Heun polynomials \cite{Ronveaux} corresponding to the truncated solutions of the Heun equation. The explicit calculation of the Heun polynomials $Q_n(x)$ (as well as of their corresponding eigenvalues $\t \lambda_n$) is possible only for small values of $N$. For large $N$ there are special algebraic methods allowing to reduce the problem of their determination to the calculation of the eigenvectors of a finite tri-diagonal matrix \cite{Ronveaux}.

Assume that the Heun polynomials are already found. The eigenfunctions of the operator $L$ are well known, they are the Jacobi polynomials $P_n(x)$. Hence we can write down the expansion
\be
Q_n(x)= \sum_{s=0}^N W_{ns} P_s(x) \lab{Q_P_W} \ee
with $W_{ns}$ some coefficients.

From \re{M_3-diag} and \re{M_Q_lam} we obtain that the coefficients $W_{ns}$ satisfy the 3-term recurrence relation
\be
W_{n,s+1} \zeta_{s+1} + W_{ns} \eta_s + W_{n,s-1} \xi_s = \t \lambda_n W_{ns}, \quad s,n =0,1,\dots, N. \lab{rec_W} \ee
From this relation it follows that one can express $W_{ns}$ in terms of orthogonal polynomials 
\be
W_{ns} = W_{n0} R_s(\t \lambda_n), \lab{W_R} \ee
where the polynomials $R_s(x)$ satisfy the recurrence relation
\be
\zeta_{s+1} R_{s+1}(x) + \eta_s R_s(x) + \xi_s R_{s-1}(x) = x R_s(x). \lab{} \ee

Hence the overlap coefficients (or the interbasis coefficients) $W_{ns}$ are expressed in terms of orthogonal polynomials $R_n(x)$. These polynomials become the Racah polynomials when $\kappa=0$ (i.e. when the algebra of the operators $L,M$ becomes the Racah algebra). It is natural therefore to call $R_n(x)$ the Racah-Heun orthogonal polynomials.

Let us consider the Racah-Heun polynomials $R_n(x)$ in more details. It is convenient to introduce the monic polynomials
\be
\hat R_n(x) = g_n R_n(x) = x^n + O(x^{n-1}) , \lab{monic_R} \ee
where $\delta_n$ is a factor defined by the condition
\be
\zeta_{n+1}\delta_{n+1} = \delta_n \lab{g_n} \ee
Then the polynomials $\hat R_n(x)$ satisfy the recurrence relation
\be
\hat R_{n+1}(x) + B_n \hat R_n(x) + U_n \hat R_{n-1}(x) = x \hat R_n(x), \lab{rec_R_monic} \ee
where the recurrence coefficients $B_n, U_n$ can be found as in  \re{M_3-diag}
\ba
&&B_n = \eta_n = (\tau_1 + \tau_2)b_n + \tau_3 b_n + \tau_4 \lambda_n  \nonumber \\
&&U_n= u_n \xi_n \zeta_{n} = u_n (\tau_1 \lambda_{n-1} + \tau_2 \lambda_n + \tau_3)(\tau_1 \lambda_{n} + \tau_2 \lambda_{n-1} + \tau_3)  \lab{BU} \ea
Introduce also the monic Wilson polynomials $W_n(x;a_1,a_2,a_3,a_4)$ satisfying the recurrence relation \cite{KLS}
\begin{equation}
 \label{rec_Wilson}
W_{n+1}(x) + (A_n +C_n - a_1^2)\,W_n(x) + A_{n-1} C_n W_{n-1}(x) = xW_n(x), \end{equation}
where
\begin{align*}
 \begin{aligned}
  A_n&=\frac{(n+g-1)(n+a_1+a_2)(n+a_1+a_3)(n+a_1+a_4)}{(2n +g-1)(2n+g)},
  \\
  C_n&=\frac{n(n+a_2+a_3-1)(n+a_2+a_4-1)(n+a_3+a_4-1)}{(2n+g-2)(2n +g-1)},
 \end{aligned}
\end{align*}
and where  $g=a_1+a_2+a_3+a_4$.

In \cite{GIVZ} it was shown that the 3-diagonalization procedure of the hypergeometric operator $L$ with $\tau_4=0$ gives the operator $M$ which generates the Wilson polynomials 
\begin{equation}
\label{Q_W_id}
Q_n(x) = W_n(\gamma-x;a_1,a_2,a_3,a_4),
\end{equation}
with $\gamma = \frac{1}{2} \left( a_1 + a_2 - a_1^2 - a_2^2 \right)$ and where
\begin{alignat*}{2}
\omega_1&=a_1+a_2-1, &\qquad \omega_2&=a_3+a_4-1,
\\
\mu_1&= 1-a_1-a_3, &\qquad \mu_2&=1-a_2-a_3.
\end{alignat*}
The parameters $\mu_1, \mu_2$ in the above formulas are related with the parameters $\tau_2,\tau_3$ by the formulas \cite{GIVZ}
\begin{equation}
\label{tau_nu}
\tau_2= \frac{1}{2} (1+\mu_1-\mu_2), \qquad \tau_3 = \mu_1\mu_2 +\frac{(\omega_1+\omega_2)(\mu_1+\mu_2-1)}{2}.
\end{equation}
(recall that the condition $\tau_1 + \tau_2=1$ is assumed).

It is thus seen that the Racah-Heun polynomials $\hat R_n(x)$ can be considered as a one-parameter perturbation of the Wilson polynomials $W_n(x)$. Indeed, the recurrence coefficient $U_n$ of the Racah-Heun polynomials coincide with the corresponding recurrence coefficient $A_n C_{n-1}$ of the Wilson polynomials. The only difference occurs for the diagonal recurrence coefficients:
\be
B_n = B_n^{(0)} +  \tau_4 \lambda_n, \lab{BB_tau} \ee
where $B_n^{(0)}$ stands for the diagonal recurrence coefficient of the Wilson polynomials. The parameter $\tau_4$ plays the role of the perturbation parameter: if $\tau_4=0$ then we return to pure Wilson polynomials.  This observation allows to find approximate expressions for the Racah-Heun polynomials if the the parameter $\tau_4$ is sufficiently small.

With the help of the Heun equation \re{red_Heun}, we can thus relate two systems of polynomials: the (nonorthogonal) Heun polynomials $Q_n(x)$ which are polynomial solutions of the Heun equation under the truncation condition \re{tr_M} and the Racah-Heun polynomials $R_n(x)$ which are orthogonal polynomials corresponding to the expansion \re{Q_P_W} of the Heun polynomials over the basis of the Jacobi polynomials. Note that the 3-diagonal property of the Heun functions with respect to the Gauss hypergeometric functions is well known \cite{Ronveaux}. This property follows easily from Racah-Heun quadratic algebra \re{Heun_alg}.

Note that the restriction  to polynomial solutions of the Heun equation is closely related to the quasi-exact solvability of certain differential equations \cite{Tur}, \cite{BGK}, \cite{LV}.

It is interesting to note that the quadratic algebra \re{Heun_alg} already appeared in the analysis of quantum superintegrable systems  \cite{Dask}, \cite{Marq}. In the next section we describe the relation of this algebra to the Racah problem for the $su(1,1)$ Lie algebra.

\section{Racah-Heun algebra and the Racah problem for the $su(1,1)$ algebra}
\setcounter{equation}{0}
There is a simple realization of the Racah-Heun algebra in terms of the intermediate Casimir operators that occur in  the Racah problem for the $su(1,1)$ algebra.

Consider the $\mathfrak{su}(1,1)$ algebra with generators $S_0, S_{+}, S_{-}$ and commutation relations 
\begin{equation*}
[S_0, S_{\pm}]= \pm S_{\pm}, \quad [S_-, S_+]=2 S_0.
\end{equation*}
The Casimir operator
\begin{equation*}
C \equiv S_0^2 - S_0 - S_+S_- ,
\end{equation*}
commutes with all generators. For the irreducible representations of $\mathfrak{su}(1,1)$ that are relevant to the present paper, the Casimir operator takes the value $\sigma(\sigma-1)$.  When $\sigma>0$, the irreducible representations of the positive-discrete series are defined on the space spanned by the basis vectors $e_{n}$, $n=0,1,\ldots$, by the following actions:
\begin{equation*}
S_0 e_n = (n+\sigma) e_n,\quad  S_+ e_n = \gamma_{n+1} e_{n+1}, \quad S_- e_n = \gamma_n e_{n-1}, 
\end{equation*} 
where $\gamma_n = \sqrt{n(2 \sigma +n-1)}$. Consider the tensor product of three representations of the positive-discrete series with representation parameters $\sigma_1$, $\sigma_2$ and  $\sigma_3$. We use the notation
\begin{align*}
 S^{(ij)}_0=S^{(i)}_0+S^{(j)}_0,\qquad S^{(ij)}_{\pm}=S^{(i)}_{\pm}+S^{(j)}_{\pm},\qquad i,j=1,2,3,
\end{align*}
where the superscript $i$ in $S^{(i)}$ specifies on which representation space the generator acts.

The Casimir operators 
\be
C_i = {{S^{(i)}_0}}^2 - {S^{(i)}_0}- {S^{(i)}_+} {S^{(i)}_-} , \quad i=1,2,3 \lab{1-mod_C} \ee
commute with all operators under consideration and can be chosen as constants 
\be
C_i = \sigma_i (\sigma_i-1) \lab{C_i_sigma} \ee
The intermediate Casimir operators $C_{ik}$ are defined as
\begin{equation*}
C_{ik} = [S_0^{(ik)}]^2 - S_0^{(ik)} - S_+^{(ik)}S_-^{(ik)},\qquad\text{for}\quad (ik)\in \{(12),(23), (13)\},
\end{equation*}
and the total Casimir operator $C$ is written as
\begin{equation*}
C=[S_0^{(123)}]^2 - S_0^{(123)} - S_+^{(123)}S_-^{(123)},
\end{equation*}
where $S_{\pm}^{(ijk)}=S^{(i)}_{\pm}+S^{(j)}_{\pm}+S^{(k)}_{\pm}$ and similarly for $S_0^{(ijk)}$. By construction, the total Casimir operator commutes with all intermediate Casimir operators. 

Moreover, there is a simple relation between all Casimir operators \cite{GVZ_equi} 
\be
C = C_{12} + C_{23} + C_{31} -C_1 - C_2 - C_3 \lab{total_rel}. \ee

Consider the decomposition of the three-fold tensor product representation in irreducible components. On each of these components, the total Casimir operator takes the value $C=\sigma_4(\sigma_4-1)$, where 
\be
\sigma_4=N+\sigma_1+\sigma_2+\sigma_3 \lab{sigma_res} \ee
and $N$ is a non-negative integer. For a given $N$, there corresponds $N+1$ eigenvalues of the intermediate Casimir operators $C_{ij}$ which are of the form $\sigma_{ij}(\sigma_{ij}-1)$ where $\sigma_{ij}=k_{ij}+\sigma_i+\sigma_j$ and where $k_{ij}=0,1,\ldots,N$. Let $e_{k_{12}}^{(12)}$ and $e_{k_{23}}^{(23)}$ be eigenbases for $C_{12}$ and $C_{23}$, i.e.
\begin{equation*}
C_{12} e_{k_{12}}^{(12)} = \sigma_{12} (\sigma_{12} -1) e_{k_{12}}^{(12)}, \quad C_{23} e_{k_{23}}^{(23)} = \sigma_{23} (\sigma_{23} -1) e_{k_{23}}^{(23)},
\end{equation*}
where $ k_{12}, k_{23}=0,1,\dots,N$. The Racah problem consists in obtaining the expansion coefficients of the basis $e_{k_{23}}^{(23)}$ over the basis $e_{k_{12}}^{(12)}$
\begin{equation*}
e_{k_{23}}^{(23)} = \sum_{k_{12}=0}^N R_{k_{12}k_{23}} e_{k_{12}}^{(12)}.
\end{equation*}
A one-dimensional model for the Racah problem was constructed in \cite{GVZ_equi}  by separation of variables and dimensional reduction. It allows to express the three intermediate Casimir operators as second-order differential operators in one variable. One has
\begin{align}
\label{S}
\begin{aligned}
 C_{12}&= x^2(1-x)\, \partial_x^2 + x[(N-1-2 \sigma_1)x + 2(\sigma_1 + \sigma_2)]\partial_x
 \\
 &\qquad \qquad +2N \sigma_1 x +(\sigma_1+\sigma_2)(\sigma_1+\sigma_2-1),
 \\[.2cm]
 C_{23}&= x(x-1)\, \partial_x^2 + [2(1-N-\sigma_2 -\sigma_3)x + (N-1+2\sigma_3)]\partial_x 
 \\
 &\qquad \qquad +(N+\sigma_3+\sigma_2)(N+\sigma_3+\sigma_2-1),
 \\[.2cm]
 C_{31}&=x(x-1)^2 \partial_x^2 + (1-x)[(N-1-2\sigma_1)x + 1-N-2\sigma_3]\partial_x 
 \\
 &\qquad \qquad +2N \nu_1 (1-x) + (\sigma_3+\sigma_1)(\sigma_3+\sigma_1-1) .
 \end{aligned}
\end{align}
Now it is easily seen that (up to an affine transformation) the operator $C_{23}$ is the hypergeometric operator \re{L_hyp} while the operator $C_{12}$ coincide with the operator $M$ obtained from the hypergeometric operator $L$ by the tridiagonalzaition procedure with $\tau_4=0$. We thus have the Racah algebra  \re{W_def} for the differential operators $L$ and $M$. This was observed in \cite{GIVZ}. 

If we now consider the general tridiagonalization scheme \re{M_def} with $\tau_4 \ne 0$ then it is clear that the new operator $M$ differs from the previous one only from the addition of the term proportional to the operator $L=C_{23}$. We thus see that (up to affine transformations of the operators $L,M$) one can choose a realization of the Racah-Heun algebra as 
\be
L = C_{23}, \quad M = C_{12} + \beta C_{23} \lab{LM_su(1,1)} \ee
with arbitrary parameter $\beta$.

There are two exceptional cases: $\beta=0$ and $\beta=1$ when the Racah-Heun algebra degenerates to the Racah algebra. Indeed, when $\beta=0$ we return to the Racah problem with the intermediate Casimir operators  $C_{23}, C_{12}$. When $\beta=1$ we obtain the operator $M= C_{12} + C_{23}$ which due to \re{total_rel} is equal (up to an affine transformation) to the operator $C_{31}$ and we return to the Racah problem for the operators $C_{23}$ and $C_{31}$. For generic choices of the parameter $\beta$ we obtain the operator $M= C_{23}+\beta C_{12}$ which together with $C_{23}$ generate the Racah-Heun algebra. This provides a simple interpretation of the Racah-Heun algebra in terms of intermediate Casimir operators in the addition of 3 $su(1,1)$ representations.

The operators  $C_{12}$ and $C_{23}$ (in the appropriate realization) moreover appear to be integrals of motion of the most general scalar superintegrable model of order 2 on the 2-dimensional sphere \cite{KMP}, \cite{GVZ_Racah}. The Hamiltonian of this system is
\be
H = {\bf J}^2 + \sum_{i=1}^3 \frac{a_i}{x_i^2}, \lab{Ham_S9} \ee
where $\bf J$ is the angular momentum operator and the parameters of the potential are expressed in terms of the $su(1,1)$ representation parameters $\sigma_i$ as
\be
a_i = 4(\sigma_i-3/4)(\sigma_i-1/4) \lab{a_i}. \ee 
The motion is restricted to the 2-dimensional sphere $x_1^2 + x_2^2 + x_3^2=1$ and the Hamiltonian of the system is related to the total Casimir operator of $su(1,1)$ as follows
\be
H = 4 C_4 + 3/4 \lab{Ham_C}. \ee
The $su(1,1)$ intermediate Casimir operators $C_{12}, C_{23}, C_{32}$ are integrals of motion, as is obvious because they commute with the Hamiltonian \re{Ham_C}.

On the one hand, the overlap coefficients between the wave functions corresponding to the operators  $C_{12}$ and $C_{23}$ are expressed in terms of the Racah-Wilson polynomials as expected because these overlap coefficients coincide with the Racah coefficients of the $su(1,1)$ algebra. On the other hand, the diagonalization of the integrals $C_{12}$ and $C_{23}$ corresponds to the separation of variables in the model \re{Ham_S9} in two spherical coordinate systems.  The diagonalization of the operator $C_{12} + \beta C_{23}$ corresponds rather to the separation of variables in elliptic coordinates \cite{KMP}. This gives an additional interpretation of the tridiagonalization procedure of the hypergeoemtric operator which is connected with separation of variables in elliptic coordinates.

\section{Finite-dimensional representations of the Racah-Heun algebra}
\setcounter{equation}{0}
This section focuses on finite-dimensional representations of the Racah-Heun algebra \re{Heun_alg} with arbitrary real structure coefficients $\alpha_{1,2}, \gamma_{1,2}, \epsilon_{1,2} ,\delta, \kappa$. We already know that when $\kappa=0$ the Racah-Heun algebra becomes the Racah algebra. The finite-dimensional representations of the Racah algebra are described in \cite{TV} for instance (see also  \cite{GVZ_Racah},  \cite{GVZ_Racah2}). For generic values of the structure parameters, all finite-dimensional representations of the Racah algebra are described by the Leonard pair of matrices $L,M$. This means the following. There exists a basis $e_n$  where the matrix $L$ is diagonal and nondegenerate (i.e. all its eigenvalues are distinct) and the matrix $M$ is nondegenerate and symmetric 3-diagonal:
\be
L e_n = \lambda_n e_n, \quad M e_n = a_{n+1} e_{n+1} + b_n e_n + a_n e_{n-1} , \quad n=0,1,\dots, N \lab{Leon_e}. \ee
Similarly, there exists a (dual) basis $d_n$  where the matrix $M$ is diagonal while the matrix $L$ is 3-diagonal
\be
M d_n = \mu_n d_n, \quad L d_n = \xi_{n+1} d_{n+1} + \eta_n d_n + \xi_n d_{n-1} , \quad n=0,1,\dots, N \lab{Leon_d} \ee
The spectral values $\lambda_n, \: \mu_n$ are quadratic functions in $n$ while the coefficients $a_n, b_n$ coincide with the recurrence coefficients of the Racah polynomials. By "generic case", we mean that the spectra of the matrices $L$ and $M$ are not degenerate (see \cite{TV} for details).

Consider now the finite-dimensional representations of the Racah-Heun algebra \re{Heun_alg}. This means that the operators $L$ and $M$ are represented by square matrices of dimension $N+1 \times N+1$. Again it is assumed that these matrices are nondegenerate.

We can eliminate the term $\kappa L^2$ in relations \re{Heun_alg} by introducing the new generator 
\be
K=M-\rho L \lab{KML} \ee
instead of $M$, where
the parameter $\rho$ is a root of the quadratic equation 
\be
3 \alpha_2 \rho^2 + 3 \alpha_1 \rho + \kappa =0 \lab{eq_rho}. \ee
It then follows that the pair of operators $A_1=L$ and $A_2=K$ will realize the Racah algebra \re{W_def} with structure parameters depending on $\kappa$. We have thus brought finite-dimensional representations of the Racah-Heun algebra in the framework of the Racah algebra. This means that there exists a basis $e_n$ where the operator $L$ is diagonal while the operator $M = K +\rho L$ is 3-diagonal as in \re{Leon_e}. 

In this basis the eigenvalue $\lambda_n$ of the operator $L$ is a quadratic polynomial in $n$.  The coefficient $a_n$ is basically the same as for the Racah polynomials. The diagonal coefficient $b_n$ differs from the corresponding diagonal coefficient $b_n^{(0)}$ of the Racah polynomials by an additional term:
\be
b_n = b_n^{(0)} + \rho \lambda_n \lab{b_RH} \ee   
as follows from\re{KML}.

However for the dual basis $d_n$ where the operator $M$ is diagonal $M d_n = \mu_n d_n$, the operator $L$ is not 3-diagonal. Indeed, write
\be
L d_n = \sum_{m=0}^N L_{mn} e_m \lab{Ldn} \ee
with some matrix $L_{mn}$. Substituting \re{Ldn} into \re{str_RH}, we find that the matrix  $L_{mn}$ cannot be tridiagonal. Moreover, the eigenvalues $\mu_n$ of the operator $M$ cannot be found in an explicit analytical  form for arbitrary $N$. This leads to the conclusion that  the Racah-Heun algebra does not belong to the class of  algebras encompassing Leonard pairs. 

Note that from equation \re{eq_rho} it follows that real solutions for $\rho$ are possible only if 
\be
3 \alpha_1^2 -4 \alpha_2 \kappa >0 \lab{cond_kap} \ee
(condition \re{cond_kap} holds, e.g. for sufficiently small values of $\kappa$). In this case the finite-dimensional representations of the Racah-Heun algebra are equivalent to the Racah problem for two intermediate Casimir operators  \re{LM_su(1,1)}.

Otherwise the parameter $\rho$ is complex and the operator $K$ cannot be self-adjoint. Describing finite-dimensional representations of the Racah-Heun algebra in this case is an interesting open problem.

\section{Tridiagonalization of the Laguerre, Hermite and free motion operators}
\setcounter{equation}{0}
Consider now the Laguerre operator
\be
L = x \partial_x^2 + (a+1-x) \partial_x  \lab{Lag_L} \ee
which is the confluent hypergeometric operator. 

Applying the same Ansatz as in the previous section, we construct the operator $M$ according to \re{M_def} and assume the same normalization condition $\tau_1+\tau_2=1$
\ba
&&M =  x(x+\tau_4) \partial_x^2 + \left(-x^2 + (3+a -\tau_4-2 \tau_1)x + \tau_4 (1+a) \right) \partial_x + \nonumber \\
&&(\tau_3-\tau_2)x +1-\tau_1 + a\tau_2. \lab{M_Lag} \ea
It can be showed that the eigenvalue equation $M \psi(x) = \lambda \psi(x)$ becomes the confluent Heun equation \cite{Ronveaux}. 

The algebra constructed from the operators $L,M$ and $Z=[L,M]$ has similar commutation relations \re{Heun_alg} with the following structure constants
\ba
&&\alpha_1=-2, \; \alpha_2=0 , \; \kappa = 6 \tau_4, \; \delta= a+1 -2 \tau_3 + \tau_4, \nonumber \\
&&\gamma_1 = 4\tau_1 \tau_2 + \tau_4 (4 \tau_3 - \tau_4 -2a-2), \; \gamma_2 =-1, \nonumber \\
&& \ve_1 = -\tau_3 a^2 + a(2\tau_1 \tau_2 - \tau_3 \tau_4) + \tau_3(1-\tau_4) + 2\tau_1 \tau_2, \; \ve_2= (a+1) \tau_3. \lab{str_Lag} \ea
When $\tau_4=0$, we have $\kappa=0$ and the resulting algebra becomes the Hahn algebra. When $\tau_4 \ne 0$, we have a generalization of the Hahn algebra which can be called the Hahn-Heun algebra. All the considerations of the previous section can be translated to this case without difficulties.

Finally, consider the Hermite operator
\be
L= \partial_x^2 -2 x  \partial_x \lab{Her_L}. \ee
The operator $M$ is (again with condition $\tau_1 + \tau_2=1$  assumed)
\ba
&&M = \tau_1 X L + \tau_2 L X + \tau_3 X + \tau_4 L = \nonumber \\
&&(x+\tau_4) \partial_x^2 + 2(\tau_2 -\tau_4 x -x^2) \partial_x +(\tau_3 -2 \tau_2) x. \lab{Her_M} \ea
The eigenvalue equation $M \psi(x) = \lambda \psi(x)$ becomes the double confluent Heun equation \cite{Ronveaux}.

The corresponding quadratic algebra \re{Heun_alg} has now the structure constants
\ba
&&\alpha_1=\alpha_2=0, \; \kappa=-6, \; \gamma_1=4(1-2 \tau_3 - \tau_4^2), \nonumber \\
&&\gamma_2 = -4, \; \ve_1 = 2(2-\tau_3) - 8 \tau_1 \tau_2, \; \ve_2=0 \lab{str_Her}. \ea 
Note that when $\alpha_1=\alpha_2=0$,  the Racah algebra \re{W_def} becomes the Lie algebra isomorphic to $sl_2$ (or a degenerate case e.g. the Heisenberg-Weyl algebra). Hence, the quadratic algebra with structure constants \re{str_Her} can be considered to be the "minimal" quadratic extension of the linear $sl_2$ algebra.

\section{Conclusion}

Summing up, we have seen that the Heun operator results from the most general tridiagonalization of the hypergeometric operator. It has been found that these two operators generate the Racah-Heun algebra - a one parameter extension of the Racah algebra. This Racah-Heun algebra provides a natural interpretation of the fact that the expansion coefficients of the Heun functions in terms of hypergeometric functions obey 3 term recurrence relations. Under a truncation condition, this has allowed to define the Racah-Heun orthogonal polynomials that arise in the expression of the non-orthogonal Heun polynomials in terms of the Jacobi polynomials; these Racah-Heun polynomials thus form a representation space for the Racah-Heun algebra. The connection to quasi-exact solvability was pointed out. It was also indicated how the Racah-Heun algebra can be realized by generators constructed from the intermediate Casimir operators arising in the coupling of 3 $su(1,1)$ representations. From the connection that the $su(1,1)$ Racah problem has with superintegrable models, we have been able to relate ithe Racah-Heun algebra to separation of variables in elliptic coordinates which is known to be a context in which the Heun equation occurs. The confluent and double confluent Heun equation have been similarly given an algebraic interpretation.

We recall in concluding that this study was undertaken with time and band limiting in mind. A forthcoming publication \cite{GVZband} will indicate how the tridiagonalization ideas are instrumental in the construction of the differential operators that commute  with localization integral operators.

\bigskip\bigskip
{\Large\bf Acknowledgments}

F. A. Gr\"unbaum  and A. Zhedanov wish to thank the Centre de Recherches Mathematiques (CRM) for its hospitality while this project was being pursued.The work of L. Vinet is supported in part by the Natural Sciences and Engineering Research Council (NSERC) of Canada.

\bigskip

\newpage

\bb{99}



\bi{BGK} Y. Brihaye, S. Giller and P. Kosinski, {\it Heun equations and quasi exact solubility}, J.Phys.A: Math and Gen., {\bf 28} (1995), 421--431.

\bi{Chi} T. Chihara, {\it An Introduction to Orthogonal
Polynomials}, Gordon and Breach, NY, 1978.

\bi{Dask} C.Daskaloyannis, {\it Quadratic Poisson algebras of two-dimensional classical superintegrable systems and quadratic associative algebras of quantum superintegrable systems}, J. Math. Phys. {\bf 42}, 1100--119 (2001). arXiv:math-ph/0003017v1.

\bi{DG} J. J. Duistermaat and F. A. Gr\"unbaum, {\it Differential Equations in the Spectral Parameter}, Commun. Math. Phys. {\bf 103}, (1986) 177-240

\bi{Erd} A. Erd\'elyi, {\it Certain expansions of solutions of the Heun equation}, Quart J. Math. Oxford
Ser. (2), {\bf 15} (1944), 62--69.

\bi{G} F. A. Gr\"unbaum, {\it Band-time-band limiting integral operators and commuting differential operators}, Algebra and Analysis {\bf 8}, (1996) 122-126

\bi{GVZband}F. A. Gr\"unbaum, L. Vinet and A. Zhedanov, {\it Space and energy limiting in quantum mechanics}, in preparation

\bi{GVZ_equi} V.Genest, L.Vinet and A.Zhedanov, The equitable Racah algebra from three $su(1,1)$ algebras {\it} J. Phys. A: Math. Theor. 47 (2014) 025203.  arXiv:1309.3540v2.

\bi{GVZ_Racah} V.Genest, L.Vinet and A.Zhedanov, {\it Superintegrability in Two Dimensions and the RacahWilson Algebra}, Lett.Math.Phys. {\bf 104}, 931--952 (2014). arXiv:1307.5539v1.
\bi{GVZ_Racah2} V.Genest, L.Vinet and A.Zhedanov, {\it The Racah algebra and superintegrable models}, J. Phys.:Conf. Ser. {\bf 512} (2014) 012011. arXiv:1312.3874.
\bi{GIVZ} V. Genest, M.E.H. Ismail, L. Vinet, A. Zhedanov {\it Tridiagonalization of the hypergeometric operator and the Racah-Wilson algebra}, arXiv:1506.07803.

\bi{GVZ} V.Genest, L.Vinet and A.Zhedanov, {\it The non-symmetric Wilson polynomials are the Bannai-Ito polynomials}, arXiv:1507.02995

\bi{GZ_6j}  Ya. A. Granovskii and A. Zhedanov. {\it Nature of the symmetry
group of the 6j-symbol}. Zh. Eksp. Teor. Fiz., {\bf 94}:49-54, 1988.

\bi{GZ_preprint} Ya. A. Granovskii and A. Zhedanov. {\it Exactly solvable
problems and their quadratic algebras}. Preprint DONFTI-89-7, 1989.

\bi{GLZ_Annals} Ya. A. Granovskii, I.M. Lutzenko, and A. Zhedanov.
{\it Mutual integrability, quadratic algebras, and dynamical symmetry}. Ann. Phys., {\bf 217}: 1-20, 1992.

\bi{In}E. L. Ince, {\it Ordinary differential equations}, Dover, New York, 1944

\bi{Ismail} M.E.H.Ismail, {\it Classical and Quantum orthogonal polynomials in one variable}.
Encyclopedia of Mathematics and its Applications (No. 98), Cambridge, 2005.

\bi{IK1} M. E. H. Ismail and E. Koelink. {\it Spectral analysis of certain Schr\"odinger operators}. SIGMA,
{\bf 8}: 61-79, 2012.

\bi{IK2} M. E. H. Ismail and E. Koelink. {\it The J-matrix method}.
Adv.Appl.Math., {\bf 56}, 379--395, 2011.

\bi{KM}  E.G.Kalnins and W.Miller, Jr., {\it Hypergeometric expansions of Heun polynomials}, SIAM J. Math. Anal.
{\bf 22}, 1450--1459.

\bi{KMP} E.G.Kalnins, W.Miller and S.Post, {\it Wilson polynomials and the generic
superintegrable system on the 2-sphere}, J. Phys. A: Math. Theor., {\bf 40}:1152511538, 2007.

\bibitem{KLS} R. Koekoek, P.A. Lesky, and R.F. Swarttouw. {\it Hypergeometric orthogonal polynomials and their q-analogues}. Springer, 1-st edition, 2010.

\bi{LV} P.L\'etourneau and L.Vinet,  {\it Superintegrable Systems: Polynomial Algebras and Quasi-Exactly Solvable Hamiltonians}, Annals of Physics, {\bf 243}, (1995), 144-168.

\bi{Marq} Ian Marquette, {\it Quartic Poisson algebras and quartic associative algebras and realizations as deformed oscillator algebras}, J. Math. Phys. {\bf 54}, 071702 (2013)

\bi{Me} M. L. Mehta, {\it Random Matrices}, 2nd ed., Academic Press, San Diego, 1991

\bibitem{Ronveaux} A. Ronveaux (Ed.), {\it Heun's Differential Equations}, Oxford University Press, Oxford, 1995.

\bi{SlPo} D. Slepian and H.O. Pollak, {\it Prolate spheroidal wave functions, Fourier analysis and uncertainty, I}, Bell Syst. Tech. J., {\bf 40}, (1961) 43-64

\bi{LaPo} H. J. Landau and H. O. Pollak, {\it Prolate spheroidal wave functions, Fourier analysis and uncertainty, II}, Bell Syst. Tech. J., {\bf 40}, (1961) 65-84
\bi{LaPo2} H. J. Landau and H. O. Pollak, {\it Prolate spheroidal wave functions, Fourier analysis and uncertainty, III}, Bell Syst. Tech. J., {\bf 41}, (1962) 1295-1336
\bi{Sl2} D. Slepian, {\it On bandwith}, Proc. IEEE, {\bf 64}, (1976) 292-300
\bi{Sl3} D. Slepian, {\it Some comments on Fourier analysis, uncertainty and modeling}, SIAM. Rev., {\bf 64}, (1983) 379-393

\bi{Ter} P.Terwilliger, {\it Two linear transformations each tridiagonal with respect to an eigenbasis of the other}, Lin.Alg.Appl. {\bf 330} (2001), 149-203.

\bi{TV} P.Terwilliger and R.Vidunas, {\it Leonard pairs and the Askey-Wilson relations}, J. Algebra Appl. {\bf 03}, 411 (2004). arXiv:math/0305356.
\bi{TrW1} C. A. Tracy and H. Widom, {\it Level-Spacing Distributions and the Airy Kernel}, Commun. Math. Phys. {\bf 159} (1994), 151-174
\bi{TrW2} C. A. Tracy and H. Widom, {\it Level Spacing Distributions and the Bessel Kernel}, Commun. Math. Phys. {\bf 161} (1994), 289-309

\bi{Tur} A.Turbiner, {\it Lie algebras and polynomials in one variable}, J.Phys. A: Math and Gen., {\bf 25} (1992), L1087--L1093.

\eb

\end{document}